\documentclass[letter]{aa} % for the letters 
\def\lsim{\mathrel{\raise .4ex\hbox{\rlap{$<$}\lower 1.2ex\hbox{$\sim$}}}}
\def\gsim{\mathrel{\raise .4ex\hbox{\rlap{$>$}\lower 1.2ex\hbox{$\sim$}}}}
\usepackage{graphicx}
\usepackage{natbib}
\usepackage{txfonts}
\usepackage[breaklinks,colorlinks,citecolor=blue]{hyperref}
\begin{document}

   \title{The Local Bubble: a magnetic veil to our Galaxy}

   \author{M.I.R. Alves \inst{1,2}\thanks{Email: m.alves@astro.ru.nl},
     F. Boulanger \inst{3},
     K. Ferri\`{e}re \inst{2}
     \and
     L. Montier \inst{2}
          }

   \institute{Department of Astrophysics/IMAPP, Radboud University, PO
     Box 9010, NL-6500 GL Nijmegen, the Netherlands
   \and
  Institut de Recherche en Astrophysique et Plan\'{e}tologie (IRAP),
  CNRS, Universit\'{e} de Toulouse, CNRS, 9 Avenue du Colonel Roche,
  BP 44346, 31028, Toulouse Cedex 4, France
     \and
    \'{E}cole normale sup\'{e}rieure/LERMA, Observatoire de Paris, Sorbonne Universit\'{e}, Universit\'{e} PSL, CNRS, Paris, France 
            }

   \date{Received January 14, 2018; accepted March 06, 2018}

% \abstract{}{}{}{}{} 
% 5 {} token are mandatory

 \abstract {The magnetic field in the local interstellar medium does not follow the large-scale Galactic magnetic field. The local magnetic field has probably been distorted by the Local Bubble, a cavity of hot ionized gas extending all around the Sun and surrounded by a shell of cold neutral gas and dust. However, so far no conclusive association between the local magnetic field and the Local Bubble has been established.
 Here we develop an analytical model for the magnetic field in the shell of the Local Bubble, which we represent as an inclined spheroid, off-centred from the Sun. We fit the model to Planck dust polarized emission observations within $30^{\circ}$ of the Galactic poles. We find a solution that is consistent with a highly deformed magnetic field, with significantly different directions towards the north and south Galactic poles.
 This work sets a methodological framework for modelling the three-dimensional (3D) structure of the magnetic field in the local interstellar medium, which is a most awaited input for large-scale Galactic magnetic field models. }

   \keywords{ISM: bubbles --
                ISM: magnetic fields --
                ISM: individual objects: (Local Bubble)
               }

\titlerunning{The magnetic field in the Local Bubble}
\authorrunning{M.I.R Alves et al.}

   \maketitle
%
%-------------------------------------------------------------------
%-------------------------------------------------------------------

\section{Introduction}
\label{intro}

The interstellar medium (ISM) of our Galaxy is threaded by a pervasive magnetic field that plays an important role in many astrophysical processes  (see e.g. \citealt{Ferriere:2001}, \citealt{Heiles&Haverkorn:2012}). 
Various models of the large-scale Galactic magnetic field have been developed throughout the years (e.g. \citealt{Han:2006},  \citealt{Sun:2010}, \citealt{Jansson:2012}, \citealt{Jaffe:2013}, \citealt{Terral:2017}). These aim at describing the large-scale structure of the magnetic field, commonly divided into disk and halo components. The models are mostly constrained by observations of synchrotron emission and Faraday rotation. 
However, the observations include several local 
structures, which appear large on the sky because of their proximity, but are not part of the large-scale magnetic field.
The most important such structure is the Local Bubble,
an interstellar bubble located around the Sun and extending out to $\sim 60$\,pc towards the Galactic centre and $\gtrsim 250$\,pc towards the north and south Galactic poles \citep{Puspitarini:2014}.
The Local Bubble is thought to be the result of supernova explosions, which swept out a cavity of hot ionized gas and pushed most of the evacuated matter, together with the frozen-in magnetic field, into a dense shell of cold neutral gas and dust \citep{Cox:1987,Shelton:1998,Fuchs:2006, Lallement:2014}. 

Several observational studies have shown that the magnetic field in the local ISM does not follow the large-scale Galactic magnetic field (e.g. \citealt{Heiles:1998}, \citealt{Leroy:1999}, \citealt{Santos:2011}, \citealt{Frisch:2012}, \citealt{Berdyugin:2014}), which points approximately towards $(\ell, b) = (83^{\circ}, 0^{\circ})$ at the Sun's position 
\citep{Heiles:1996}. 
Here, $\ell$ and $b$ denote Galactic longitude and latitude, respectively.
These departures from the large-scale magnetic field 
were related to the Local Bubble shell by
\citet{Leroy:1999}, who, based on stellar polarization observations, found coherent magnetic field orientations over the distance range  60 -- 150\,pc. 
Recently, \citet{PIPXLIV:2016} analysed the Planck maps of dust polarized emission towards the southern Galactic cap ($b<-60^{\circ}$), and found that these could be represented by a uniform magnetic field 
pointing in the direction $(\ell, b) = (70^{\circ}, 24^{\circ})$, thus with a significant vertical component. At these high Galactic latitudes, the observed dust emission is most likely associated with the shell of the Local Bubble \citep{Kos:2014}. 
Despite these different studies, no definitive relation between the deformed local magnetic field and the Local Bubble has so far been established. 

In this letter, we present the first model of the magnetic field in the Local Bubble, based on a new analytical solution and on Planck observations of dust polarized emission. 
This work is a stepping stone towards modelling the 3D structure of the magnetic field in the Solar neighbourhood. 
Such a model will contribute to our understanding of the large-scale Galactic magnetic field and will also lead to a more accurate modelling of the Galactic polarized foreground emission, which contaminates observations of the $B$ modes of the cosmic microwave background \citep{BicepPlanck:2015} and of H{\sc i} emission from the epoch of reionization \citep{Jelic:2010}.

%--------------------------------------------------------------------
%--------------------------------------------------------------------
\section{The magnetic field in the Local Bubble}
\label{sec1}

Observations of dust polarized emission trace the orientation of the magnetic field averaged along the line of sight, without providing any information on the magnetic field strength. Hence, we do not attempt to model variations in the magnetic field along the line of sight, nor are we concerned with the magnetic field strength. 
Instead, we assume that all the swept-up matter and field lines are squeezed into a thin shell (whose actual thickness is irrelevant) that follows the surface of the Local Bubble.
Accordingly, the swept-up magnetic field is required to be tangent throughout to the surface of the Local Bubble. To fully define the magnetic field orientation, we make the additional assumption that the expansion motions that created the Local Bubble are purely radial from a single explosion centre.
However, these motions are not required to have spherical symmetry.

In view of the above assumptions, we now derive an analytical expression for the magnetic field, $\vec{B}$, in a very thin shell surrounding a bubble of arbitrary shape (see Appendix \ref{app} for more details).
Consider an initially homogeneous medium with uniform magnetic field $\vec{B}_{0}$,
and imagine that supernova explosions occurred at a point $O$ of this medium.
Let us define a cartesian coordinate system $(x,y,z)$ centered on $O$,
and let us denote the associated spherical coordinates by $(r,\theta,\phi)$.
Since the expansion motions driven by the explosion are assumed to be purely radial
(i.e. along the unit vector $\vec{e}_{r}$),
the initial position of a particle presently at position
$\vec{r} = r \, \vec{e}_{r}$ can be written as
\begin{equation}
\label{eq_AppA_r0}
\vec{r}_{0} = r_{0}(r,\theta,\phi) \, \vec{e}_{r} .
\end{equation}

To derive the general expression of the magnetic field deformed by the explosion,
we use the vector potential, $\vec{A}$,
defined such that $\vec{B} = \nabla \times \vec{A}$.
In the frozen-in approximation, the present vector potential, $\vec{A}$,
is related to the initial vector potential, $\vec{A}_{0}$, through
\begin{equation}
\label{eq_AppA_A}
\vec{A}(\vec{r}) = (\nabla \vec{r}_{0}) \cdot  \vec{A}_{0}(\vec{r}_{0}) .
\end{equation}
Since the initial magnetic field is uniform, we can adopt
\begin{equation}
\label{eq_AppA_A0}
\vec{A}_{0} = B_{0y}\,z\,\vec{e}_{x} + B_{0z}\,x\,\vec{e}_{y}+B_{0x}\,y\,\vec{e}_{z} .
\end{equation}
Introducing Eqs.~(\ref{eq_AppA_r0}) and (\ref{eq_AppA_A0}) into Eq.~(\ref{eq_AppA_A})
and taking $\vec{B} = \nabla \times \vec{A}$, we obtain a lengthy and unwieldy
equation for $\vec{B}$. 
However, 
under our assumption that all the field lines swept up by the explosion
are confined into a very thin shell,  
the general expression of $\vec{B}$ can be greatly simplified to
\begin{equation}
\label{eq_AppA_B_1}
\vec{B}(\vec{r}) = 
\frac{r_{0}}{r} \, \frac{\partial r_{0}}{\partial r} \ 
\left[ \vec{B}_{0t} 
+ \left( \vec{\nabla}_t R_{\rm sh} \cdot \vec{B}_{0t} \right) \, \vec{e}_{r} 
\right] ,
\end{equation}
where $R_{\rm sh}$ is the shell radius (a function of $(\theta,\phi)$),
$\vec{B}_{0t} = B_{0\theta}\,\vec{e}_{\theta} + B_{0\phi}\,\vec{e}_{\phi}$
is the orthoradial (or transverse) component of the initial magnetic field,
and $\vec{\nabla}_t = (1/r)\,(\partial/\partial \theta)\,\vec{e}_{\theta} 
+ (1/r \sin\theta)\,(\partial/\partial \phi)\,\vec{e}_{\phi}$
is the orthoradial component of the gradient operator.
The factor $(r_{0}/r)$ in Eq.~(\ref{eq_AppA_B_1}) arises from spherical divergence. 
For a very thin shell, the second factor is very large,
but this is irrelevant here because we are only interested in the direction of the magnetic field, not in its strength.

For numerical convenience, we rewrite Eq.~(\ref{eq_AppA_B_1}) in the form
\begin{equation}
\label{eq_AppA_B_2}
\vec{B}(\vec{r}) = 
\frac{r_{0}}{r} \, \frac{\partial r_{0}}{\partial r} \ 
\left[ \vec{B}_{0t} 
- \frac{\vec{n} \!\cdot\! \vec{B}_{0t}}{\vec{n} \!\cdot\! \vec{e}_{r}} \, \vec{e}_{r}
\right] ,
\end{equation}
where $\vec{n} = (\vec{e}_{r} - \vec{\nabla}_t R_{\rm sh}) / 
||\vec{e}_{r} - \vec{\nabla}_t R_{\rm sh}||$ 
is the unit vector normal to the surface of the shell.
The advantage of writing $\vec{B}$ in terms of the unit vector $\vec{n}$ is that, in contrast to $\vec{\nabla}_t R_{\rm sh}$, $\vec{n}$ is independent of the reference frame.
Interestingly, Eq.~(\ref{eq_AppA_B_2}) can also be written as
\begin{equation}
\label{eq_AppA_B_3}
\vec{B}(\vec{r}) = 
\frac{r_{0}}{r} \, \frac{\partial r_{0}}{\partial r} \, 
\frac{1}{\vec{n} \!\cdot\! \vec{e}_{r}} \ 
\left[ \vec{n} \times \left(\vec{B}_{0} \times \vec{e}_{r} \right)
\right] ,
\end{equation}
which expresses the fact that the swept-up magnetic field
is both tangent to the shell surface (as required)
and contained in the plane of the initial magnetic field, $\vec{B}_{0}$,
and the direction of the expansion motion, $\vec{e}_{r}$ (as expected).
Thus the direction of $\vec{B}$ depends only on the direction of $\vec{B}_{0}$ and on the shape of the bubble.

As we aim to model the ordered magnetic field in the shell of the Local Bubble, we do not attempt to reproduce its complex structure  \citep{Capitanio:2017, Liu:2017}. Rather, we consider that its large-scale shape can be captured by a generic geometrical form, for which we adopt a simple ellipsoid centred on the explosion centre.
The latter is not required to coincide with the Sun's position.
For a given initial magnetic field and a given bubble shape, the magnetic field in the shell has a well-determined configuration, but our view of its projection onto the plane of the sky depends on the position of the explosion centre relative to the Sun.
In practice, to fit our magnetic field model to the Planck observations, we need to perform a change of coordinates from the explosion centre frame, $(x,y,z)$, where our model equations are written, to the Sun's frame, $(x',y',z')$, where the observations are made. 
Here, we orient these two frames such that
the $x$- and $x'$-axes point towards the Galactic centre,
the $y$- and $y'$-axes point towards $\ell = 90^\circ$,
and the $z$- and $z'$-axes point towards $b = 90^\circ$.

Altogether, our magnetic field model involves 12 free parameters:
9 describing the size, the orientation, and the position of the ellipsoid,
2 defining the direction of the initial magnetic field,
and 1 relating the magnetic field orientation to the observed dust polarized emission.
More specifically, the 3 parameters giving the size of the ellipsoid are its semi-axes, $a_{\rm ell}$, $b_{\rm ell}$, and $c_{\rm ell}$.
The 3 parameters giving the orientation of the ellipsoid are the standard Euler angles, $\theta_{\rm ell}$ (nutation), $\psi_{\rm ell}$ (precession), and $\phi_{\rm ell}$ (intrinsic), defined in the explosion centre frame.
The 3 parameters giving the position of the ellipsoid are the coordinates $(\delta_x, \delta_y, \delta_z)$ of its centre in the Sun's frame.
The 2 parameters defining the direction of the initial magnetic field $\vec{B}_{0}$ are its Galactic longitude, $\ell_{0}$, and latitude, $b_{0}$.
The last parameter is the maximum dust polarization fraction, $p_{0}$, i.e. the polarization fraction obtained when the magnetic field lies exactly in the plane of the sky.

%--------------------------------------------------------------------
%--------------------------------------------------------------------
\section{Fit to Planck observations}
\label{sec2}

\begin{figure*}
\centering
\includegraphics[scale=0.36]{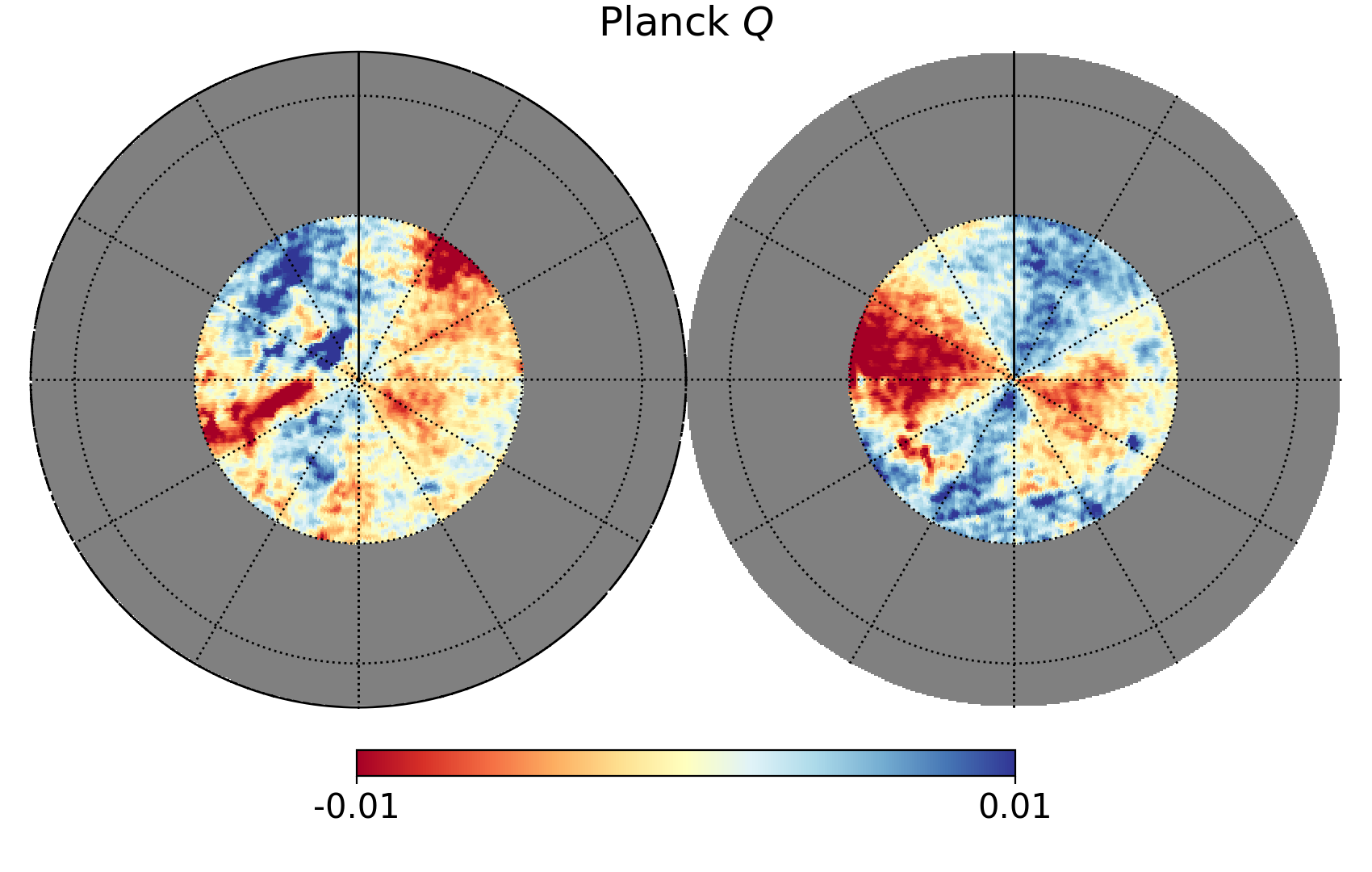}
\includegraphics[scale=0.36]{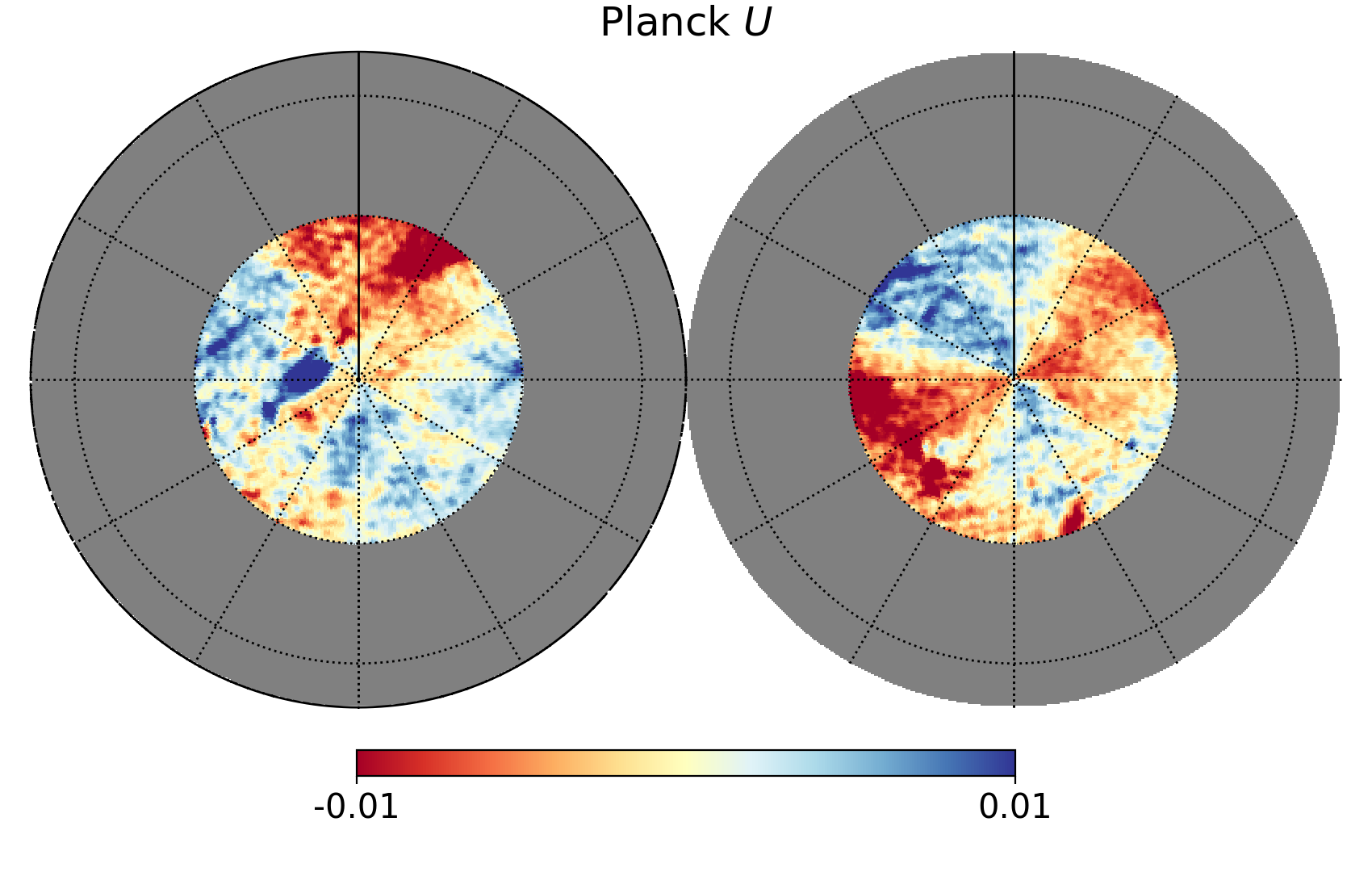}
\includegraphics[scale=0.36]{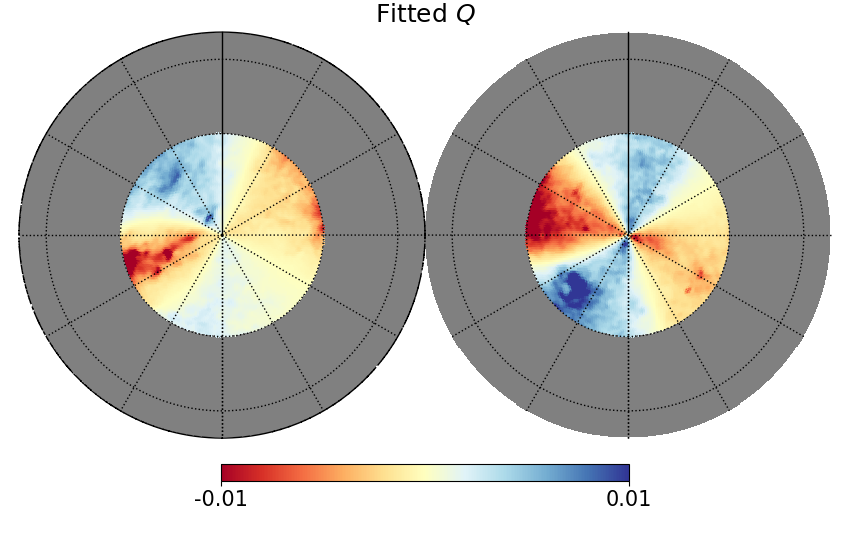}
\includegraphics[scale=0.36]{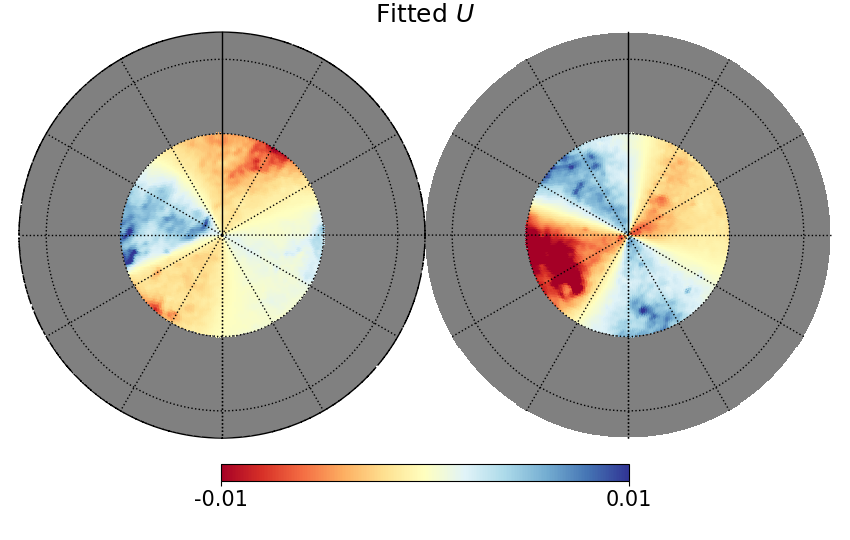}
\caption{Observed (top) and fitted (bottom) $Q$ (left) and $U$ (right) Stokes parameter maps in orthographic projection, in units of MJy\,sr$^{-1}$. Each pair of maps shows the sky divided into its two Galactic hemispheres: north (left) and south (right). The vertical full line in each map indicates $\ell=0^{\circ}$. The grey regions correspond to the areas that are masked. The texture seen in the model maps, overlaid on the regular patterns, arises from the intensity $I$. Planck observations show additional structure from the turbulent component of the magnetic field, not included in our model.}
\label{fig1}
\end{figure*}

\begin{table*}
\centering
\caption{Results of fits of our magnetic field model to Planck polarization observations. See Sect. \ref{sec1} for a description of each parameter.}
\label{tab1}
\begin{tabular}{cccccccccccc}
 $\ell_{0}$ & $b_{0}$ & $p_{0}$&  $a_{\rm ell}$ & $b_{\rm ell}$ & $c_{\rm ell}$ & $\psi_{\rm ell}$& $\theta_{\rm ell}$ & $\phi_{\rm ell}$ & $\delta_{x}$ & $\delta_{y}$ & $\delta_{z}$\\
$ \rm{[deg]}$  & [deg] & [\%] & & & & [deg] & [deg] & [deg] & & & \\
% [deg] &  [deg] &  [\%]  &  &  &  &  [deg]  &  [deg] &  [deg] &  &  &  \\
\hline\hline
$71 \pm 11 $ & $-16 \pm 7$ & $12 \pm 1 $ & 
set=1 & set=1 &  $2.7 \pm 0.6$ & $ 216 \pm 47$ & $30 \pm 50$ & set=0 & $ 0.17 \pm 0.3$  &  $0.56 \pm 0.2 $ & set=0  \\
\end{tabular}
\end{table*}

We take the latest publicly available Planck maps of the Stokes parameters $Q$ and $U$ at 353\,GHz (Planck Legacy Archive\footnote{\url{https://www.cosmos.esa.int/web/planck/pla}}, \citealt{PIP2015VIII:2016}). For the total intensity $I$ we use the model map derived by \citet{P2013XI:2014}, obtained by fitting observations that have been corrected for the cosmic microwave background anisotropies and for zodiacal light. These data are smoothed to $1^{\circ}$ angular resolution. 

Our model concerns the magnetic field in the shell of the Local Bubble only, that is, it does not include any component from the 
large-scale Galactic magnetic field. 
Therefore, to limit the contribution from the Galactic disk to the observed dust polarized emission, we restrict our analysis to the Galactic polar caps, $|b|>60^{\circ}$. 
Given this restriction, we do not attempt to fit all the parameters in the model. Instead, we present a first illustration of our method based on a limited choice for the input parameters. 
First, we neglect the vertical shift of the explosion centre relative to the Sun, that is, we set $\delta_{z}=0$, based on the argument that the Sun is located near the Galactic plane (\citealt{Joshi:2007})
and that massive stars, the most likely supernova progenitors, are also distributed close to the Galactic plane \citep{Bronfman:2000}.
Second, we consider that the shape of the Local Bubble can be described by a spheroid, and accordingly we set $a_{\rm ell}=b_{\rm ell}$. The exact value of $\phi_{\rm ell}$ is thus irrelevant, and for simplicity we adopt $\phi_{\rm ell}=0^{\circ}$. Our restriction to a spheroid follows from the observation that the Local Bubble is elongated towards the Galactic poles (see Sect. \ref{intro}), together with the limitations of our analysis, which relies exclusively on high-$|b|$ observations and is, therefore, unable to provide separate constraints on the Local Bubble dimensions in the Galactic plane. 
Finally, as dust polarized emission observations do not give any information about distances, we work with normalized sizes, with our reference length set by $a_{\rm ell}=1$. 

\begin{figure}[!ht]
\centering
\includegraphics[scale=0.15]{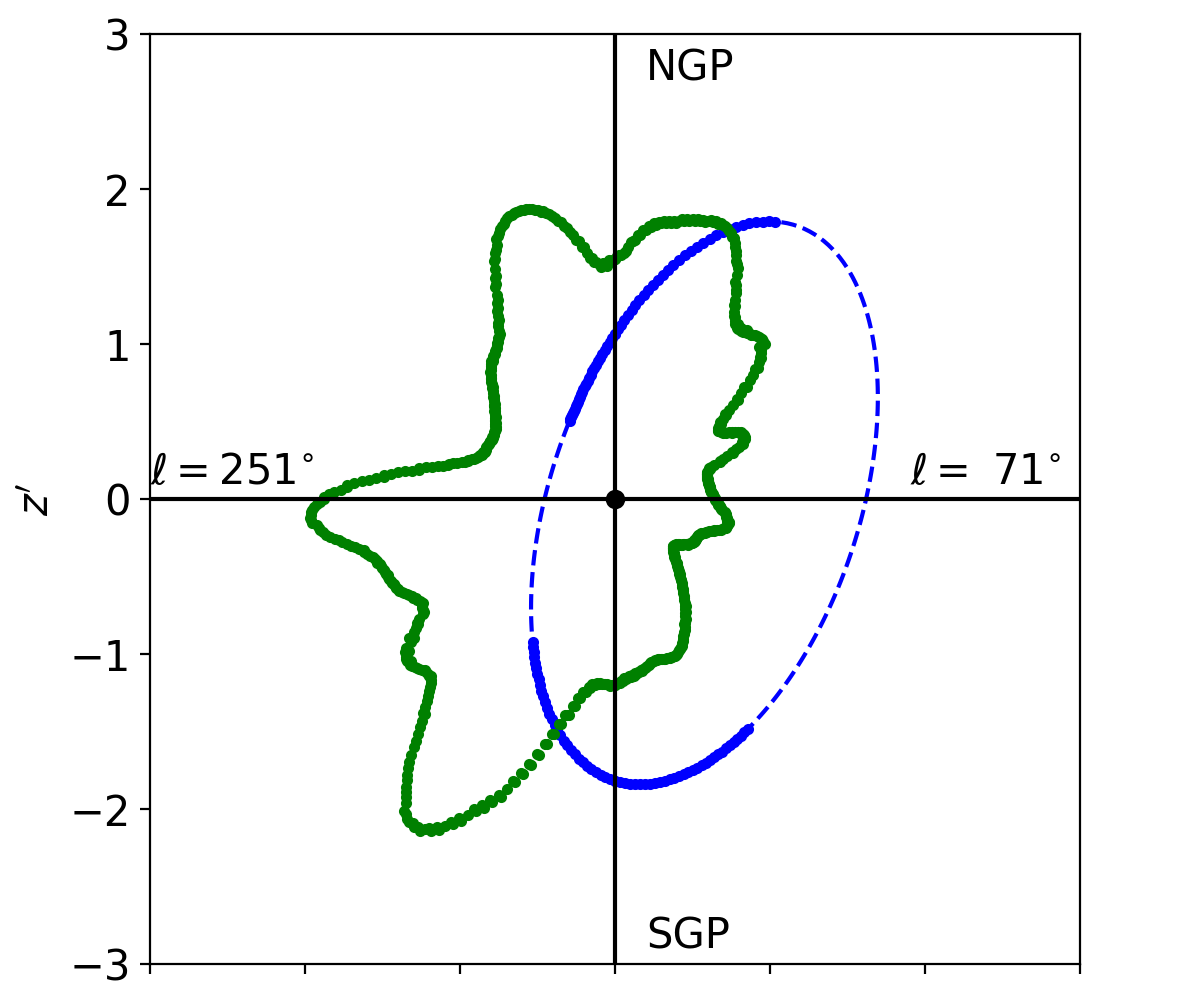}
\includegraphics[scale=0.15]{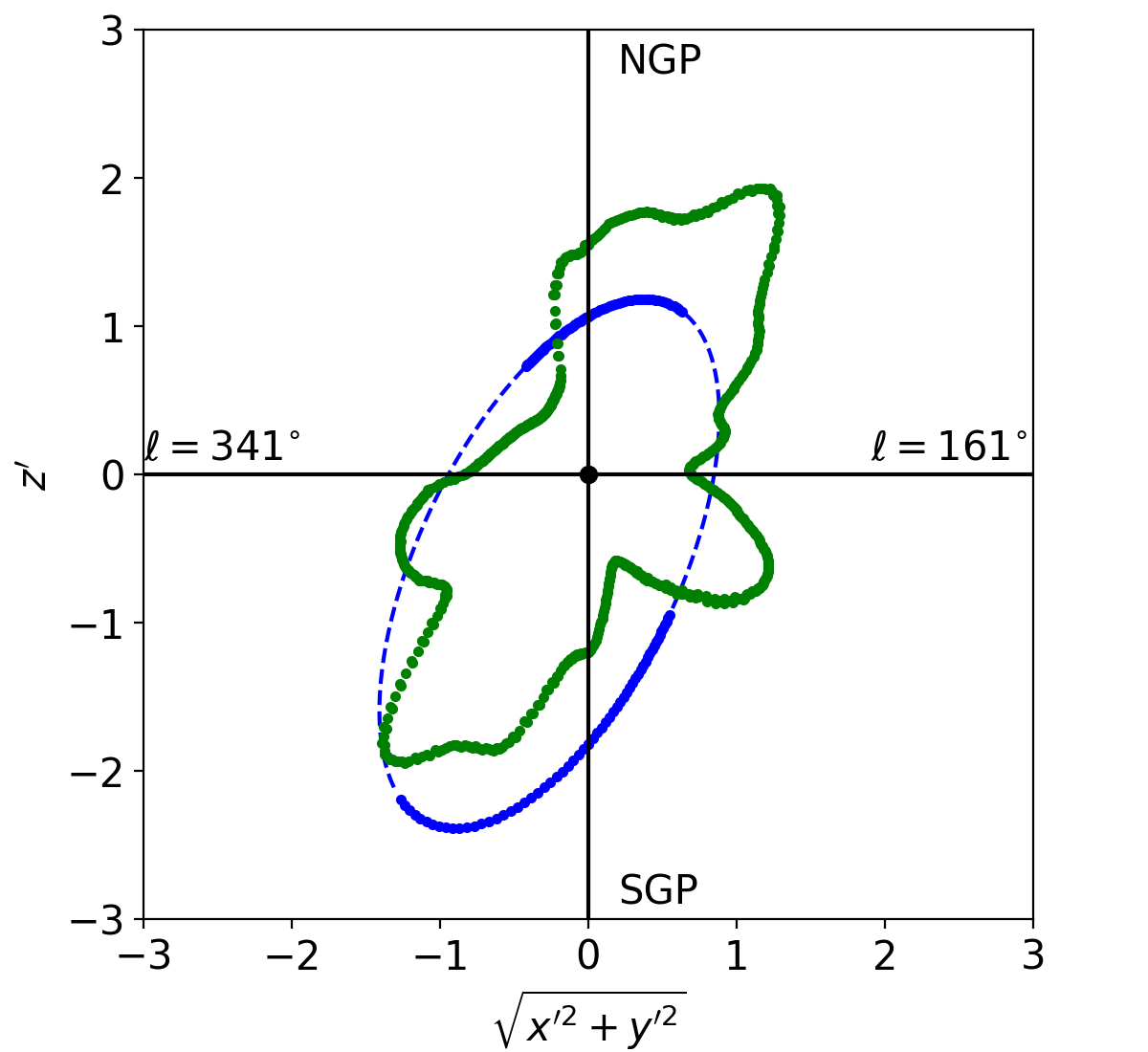}
\caption{The shape of the Local Bubble in great-circle cuts through the Galactic poles along two longitudes: $\ell_{0}$ (top) and $\ell_{0}+90^{\circ}$ (bottom). The blue curves correspond to our model and the green curves to the data from \citet{Liu:2017}. The latter are re-scaled by a factor of 1/(100\,pc). The dashed blue curves represent the sky area where $|b|< 60^{\circ}$, which is not included in the analysis.}
\label{fig2}
\end{figure}

Under the above assumptions, our magnetic field model is left with 8 free parameters out of the initial 12. We follow the methodology detailed in \citet{PIPXLIV:2016} to produce all-sky mock Stokes parameter maps from our 3D magnetic field model. In particular, we compute normalized Stokes parameters, $Q/I$ and $U/I$, which depend on the magnetic field orientation only, and then multiply them by the intensity map $I$.
We use the {\sc python mpfit} routine to derive the set of parameter values that best fit the observations. We take into account the noise in the Planck $Q$ and $U$ data, which we add in quadrature with a contribution from the turbulent magnetic field component, which is not accounted for in our model. The latter is estimated from the power spectra of the Planck dust polarization maps at 353\,GHz using data simulations from  \citet{Vansyngel:2017}. In the region of the sky under study, the mean contribution from the statistical noise is $\sigma^{\rm noise}_{Q,U}=1.4\times10^{-3}$\,MJy\,sr$^{-1}$ and that from the turbulent magnetic field is $\sigma^{\rm turb}_{Q,U}/I=0.055$. 

Table \ref{tab1} shows the results of the fit to the Planck observations, corresponding to a reduced $\chi^{2}$ of 0.8. The fitted $Q$ and $U$ Stokes parameter maps are displayed in Fig. \ref{fig1}, along with the observational Planck maps.
The uncertainties in the different parameters are derived from the standard deviations of the parameter values obtained from similar fits to 12 different Planck sub-sets of data, with and without spectral mismatch correction \citep{PlanckES:2015,PIP2015VIII:2016}, and thus account for the uncertainties due to systematic effects in the observations. The latter are significantly larger than the uncertainties resulting from the fitting procedure. 
This is particularly the case for $\psi_{\rm ell}$ and $\theta_{\rm ell}$, whose uncertainties further reflect the degeneracy between these parameters and the position of the spheroid.

We find that the initial magnetic field points in the direction $(\ell_{0}, b_{0}) = (71^{\circ}, -16^{\circ})$, which is close to that of the large-scale Galactic magnetic field at the Sun's position, $(\ell, b) = (83^{\circ}, 0^{\circ})$ (see Sect. \ref{intro}). 
The shape of the Local Bubble is consistent with a prolate spheroid of aspect ratio 2.7, whose long axis points towards $(\ell, b) = (216^{\circ}, 60^{\circ})$. 
The explosion centre is located along the direction $\ell=73^{\circ}$, similar to that of the initial magnetic field, $\ell_{0}=71^{\circ}$, 
and at a normalized distance of 0.58 from the Sun, corresponding to a large proportion (58\,\%) of the semi-minor axis of the bubble. 
In other words, the Sun is significantly off-centre. 

Our model indicates that the magnetic field points on average towards 
$(\ell, b) = (70^{\circ} \pm 11^{\circ},43^{\circ} \pm 8^{\circ})$ and $(74^{\circ} \pm 8^{\circ},-14^{\circ} \pm 18^{\circ})$ in the north and south Galactic polar caps, respectively. 
These values are obtained by averaging the magnetic field vector within $30^{\circ}$ from the poles. 
We perform the same averaging for each of the magnetic field solutions obtained from the 12 different Planck data sub-sets, and take their standard deviations as the uncertainties in $\ell$ and $b$.
The direction of the magnetic field in the southern polar cap turns out to be different from that derived by \citet{PIPXLIV:2016}, who found $(\ell, b) = (70^{\circ},24^{\circ})$, although they obtained the same maximum dust polarization fraction, $p_{0} = 12$\,\%.
The discrepancy probably finds its roots in the fundamental difference between both approaches: while \citet{PIPXLIV:2016} model the polarization data solely based on 
variations over the polar cap in the projection of a presumably uniform magnetic field,
our model includes both projection effects and intrinsic variations in the magnetic field orientation. 
Since these variations rely on a physical bubble model, our results might be more reliable. In any case, we emphasise that the results are model dependent and that our new results should be taken with particular caution, given the narrow angular range of the observations retained in our analysis added to the stringent assumptions underlying our bubble model.

%--------------------------------------------------------------------
%--------------------------------------------------------------------
\section{Discussion and conclusions}

Fitting our model of the magnetic field in the shell of the Local Bubble to Planck observations of dust polarized emission, we obtain a solution with a good reduced $\chi^{2}$. Although this solution is only indicative, it strongly suggests that the local magnetic field as probed by Planck observations is highly distorted. In particular, the directions of the magnetic field in the north and south Galactic polar caps are found to differ by $57^{\circ}$. The discrepancy between our results and those of \citet{PIPXLIV:2016}, regarding the direction of the magnetic field in the south polar cap shows that it is essential to consider physically motivated models to reproduce the observations. 

In order to effectively constrain our magnetic field model, it is necessary to extend the sky area over which the fit is performed. However, to do so we need to take into account the polarized signal arising from the Galactic disk and possibly from the halo. In principle, this can be done by combining our geometrical model of the Local Bubble with existing models of the large-scale Galactic magnetic field (e.g. \citealt{Terral:2017}). 
Further constraints come from X-ray observations and stellar astrometric data, with which the 3D structure of the Local Bubble can be estimated \citep{Capitanio:2017, Liu:2017}. 
These observations are not taken into account in the present analysis. 
As it turns out, we find that the shape of the Local Bubble derived from our fit differs from that inferred from X-ray data by \citet{Liu:2017} (as illustrated in Fig. \ref{fig2}). 
There are several reasons for this discrepancy, including the assumption by \citet{Liu:2017} that the measured X-ray intensity is simply proportional to the size of the bubble in the considered direction and our own assumption that the centre of the explosion coincides with the geometric centre of the bubble. 
When incorporating observations of the Local Bubble structure into our modelling, we will need to relax this latter assumption. 

The local magnetic field is currently not taken into account in state-of-the-art Galactic magnetic field models. However, it is undoubtedly important to consider this local contribution in order to model polarization observations outside of the Galactic plane (as recently pointed out by \citealt{PIPXLII:2016}). Here we illustrate this point using dust polarized emission observations. Clearly the same idea could be applied to other tracers of the magnetic field, including Faraday rotation measures and perhaps even synchrotron emission data.

%-------------------------------------------------------------------
%-------------------------------------------------------------------

\begin{acknowledgements}
We thank the referee for the useful comments. We also thank W. Liu for kindly providing the all-sky map of the structure of the Local Bubble. 
The results in this paper were derived using the HEALPix package \citep{Gorski:2005} and the {\sc python mpfit} routine written by M. Cappellari. This research is supported by the Agence Nationale de la Recherche (project BxB: ANR-17-CE31-0022).
\end{acknowledgements}


\bibliographystyle{aa}
\bibliography{mybib}

\begin{thebibliography}{33}
\expandafter\ifx\csname natexlab\endcsname\relax\def\natexlab#1{#1}\fi

\bibitem[{{Berdyugin} {et~al.}(2014){Berdyugin}, {Piirola}, \&
  {Teerikorpi}}]{Berdyugin:2014}
{Berdyugin}, A., {Piirola}, V., \& {Teerikorpi}, P. 2014, \aap, 561, A24

\bibitem[{{BICEP2/Keck Collaboration} {et~al.}(2015){BICEP2/Keck
  Collaboration}, {Planck Collaboration}, {Ade}, {Aghanim}, {Ahmed}, {Aikin},
  {Alexander}, {Arnaud}, {Aumont}, {Baccigalupi}, \& et~al.}]{BicepPlanck:2015}
{BICEP2/Keck Collaboration}, {Planck Collaboration}, {Ade}, P.~A.~R., {et~al.}
  2015, Physical Review Letters, 114, 101301

\bibitem[{{Bronfman} {et~al.}(2000){Bronfman}, {Casassus}, {May}, \&
  {Nyman}}]{Bronfman:2000}
{Bronfman}, L., {Casassus}, S., {May}, J., \& {Nyman}, L.-{\AA}. 2000, \aap,
  358, 521

\bibitem[{{Capitanio} {et~al.}(2017){Capitanio}, {Lallement}, {Vergely},
  {Elyajouri}, \& {Monreal-Ibero}}]{Capitanio:2017}
{Capitanio}, L., {Lallement}, R., {Vergely}, J.~L., {Elyajouri}, M., \&
  {Monreal-Ibero}, A. 2017, \aap, 606, A65

\bibitem[{{Cox} \& {Reynolds}(1987)}]{Cox:1987}
{Cox}, D.~P. \& {Reynolds}, R.~J. 1987, \araa, 25, 303

\bibitem[{{Ferri{\`e}re}(2001)}]{Ferriere:2001}
{Ferri{\`e}re}, K.~M. 2001, Reviews of Modern Physics, 73, 1031

\bibitem[{{Frisch} {et~al.}(2012){Frisch}, {Andersson}, {Berdyugin}, {Piirola},
  {DeMajistre}, {Funsten}, {Magalhaes}, {Seriacopi}, {McComas}, {Schwadron},
  {Slavin}, \& {Wiktorowicz}}]{Frisch:2012}
{Frisch}, P.~C., {Andersson}, B.-G., {Berdyugin}, A., {et~al.} 2012, \apj, 760,
  106

\bibitem[{{Fuchs} {et~al.}(2006){Fuchs}, {Breitschwerdt}, {de Avillez},
  {Dettbarn}, \& {Flynn}}]{Fuchs:2006}
{Fuchs}, B., {Breitschwerdt}, D., {de Avillez}, M.~A., {Dettbarn}, C., \&
  {Flynn}, C. 2006, \mnras, 373, 993

\bibitem[{{G{\'o}rski} {et~al.}(2005){G{\'o}rski}, {Hivon}, {Banday},
  {Wandelt}, {Hansen}, {Reinecke}, \& {Bartelmann}}]{Gorski:2005}
{G{\'o}rski}, K.~M., {Hivon}, E., {Banday}, A.~J., {et~al.} 2005, \apj, 622,
  759

\bibitem[{{Han} {et~al.}(2006){Han}, {Manchester}, {Lyne}, {Qiao}, \& {van
  Straten}}]{Han:2006}
{Han}, J.~L., {Manchester}, R.~N., {Lyne}, A.~G., {Qiao}, G.~J., \& {van
  Straten}, W. 2006, \apj, 642, 868

\bibitem[{{Heiles}(1996)}]{Heiles:1996}
{Heiles}, C. 1996, \apj, 462, 316

\bibitem[{{Heiles}(1998)}]{Heiles:1998}
{Heiles}, C. 1998, in Lecture Notes in Physics, Berlin Springer Verlag, Vol.
  506, IAU Colloq. 166: The Local Bubble and Beyond, ed. D.~{Breitschwerdt},
  M.~J. {Freyberg}, \& J.~{Truemper}, 229--238

\bibitem[{{Heiles} \& {Haverkorn}(2012)}]{Heiles&Haverkorn:2012}
{Heiles}, C. \& {Haverkorn}, M. 2012, \ssr, 166, 293

\bibitem[{{Jaffe} {et~al.}(2013){Jaffe}, {Ferri{\`e}re}, {Banday}, {Strong},
  {Orlando}, {Mac{\'{\i}}as-P{\'e}rez}, {Fauvet}, {Combet}, \&
  {Falgarone}}]{Jaffe:2013}
{Jaffe}, T.~R., {Ferri{\`e}re}, K.~M., {Banday}, A.~J., {et~al.} 2013, \mnras,
  431, 683

\bibitem[{{Jansson} \& {Farrar}(2012)}]{Jansson:2012}
{Jansson}, R. \& {Farrar}, G.~R. 2012, \apj, 757, 14

\bibitem[{{Jeli{\'c}} {et~al.}(2010){Jeli{\'c}}, {Zaroubi}, {Labropoulos},
  {Bernardi}, {de Bruyn}, \& {Koopmans}}]{Jelic:2010}
{Jeli{\'c}}, V., {Zaroubi}, S., {Labropoulos}, P., {et~al.} 2010, \mnras, 409,
  1647

\bibitem[{{Joshi}(2007)}]{Joshi:2007}
{Joshi}, Y.~C. 2007, \mnras, 378, 768

\bibitem[{{Kos} {et~al.}(2014){Kos}, {Zwitter}, {Wyse}, {Bienaym{\'e}},
  {Binney}, {Bland-Hawthorn}, {Freeman}, {Gibson}, {Gilmore}, {Grebel},
  {Helmi}, {Kordopatis}, {Munari}, {Navarro}, {Parker}, {Reid}, {Seabroke},
  {Sharma}, {Siebert}, {Siviero}, {Steinmetz}, {Watson}, \&
  {Williams}}]{Kos:2014}
{Kos}, J., {Zwitter}, T., {Wyse}, R., {et~al.} 2014, Science, 345, 791

\bibitem[{{Lallement} {et~al.}(2014){Lallement}, {Vergely}, {Valette},
  {Puspitarini}, {Eyer}, \& {Casagrande}}]{Lallement:2014}
{Lallement}, R., {Vergely}, J.-L., {Valette}, B., {et~al.} 2014, \aap, 561, A91

\bibitem[{{Leroy}(1999)}]{Leroy:1999}
{Leroy}, J.~L. 1999, \aap, 346, 955

\bibitem[{{Liu} {et~al.}(2017){Liu}, {Chiao}, {Collier}, {Cravens}, {Galeazzi},
  {Koutroumpa}, {Kuntz}, {Lallement}, {Lepri}, {McCammon}, {Morgan}, {Porter},
  {Snowden}, {Thomas}, {Uprety}, {Ursino}, \& {Walsh}}]{Liu:2017}
{Liu}, W., {Chiao}, M., {Collier}, M.~R., {et~al.} 2017, \apj, 834, 33

\bibitem[{{Parker}(1970)}]{Parker:1970}
{Parker}, E.~N. 1970, \apj, 162, 665

\bibitem[{{Planck Collaboration} {et~al.}(2014){Planck Collaboration},
  {Abergel}, {Ade}, {Aghanim}, {Alves}, {Aniano}, {Armitage-Caplan}, {Arnaud},
  {Ashdown}, {Atrio-Barandela}, \& et~al.}]{P2013XI:2014}
{Planck Collaboration}, {Abergel}, A., {Ade}, P.~A.~R., {et~al.} 2014, \aap,
  571, A11

\bibitem[{{Planck Collaboration} {et~al.}(2016{\natexlab{a}}){Planck
  Collaboration}, {Adam}, {Ade}, {Aghanim}, {Arnaud}, {Ashdown}, {Aumont},
  {Baccigalupi}, {Banday}, {Barreiro}, \& et~al.}]{PIP2015VIII:2016}
{Planck Collaboration}, {Adam}, R., {Ade}, P.~A.~R., {et~al.}
  2016{\natexlab{a}}, \aap, 594, A8

\bibitem[{{Planck Collaboration} {et~al.}(2016{\natexlab{b}}){Planck
  Collaboration}, {Adam}, {Ade}, {Alves}, {Ashdown}, {Aumont}, {Baccigalupi},
  {Banday}, {Barreiro}, {Bartolo}, {Battaner}, {Benabed}, {Benoit-L{\'e}vy},
  {Bernard}, {Bersanelli}, {Bielewicz}, {Bonavera}, {Bond}, {Borrill},
  {Bouchet}, {Boulanger}, {Bucher}, {Burigana}, {Butler}, {Calabrese},
  {Cardoso}, {Catalano}, {Chiang}, {Christensen}, {Colombo}, {Combet},
  {Couchot}, {Crill}, {Curto}, {Cuttaia}, {Danese}, {Davis}, {de Bernardis},
  {de Rosa}, {de Zotti}, {Delabrouille}, {Dickinson}, {Diego}, {Dolag},
  {Dor{\'e}}, {Ducout}, {Dupac}, {Elsner}, {En{\ss}lin}, {Eriksen},
  {Ferri{\`e}re}, {Finelli}, {Forni}, {Frailis}, {Fraisse}, {Franceschi},
  {Galeotta}, {Ganga}, {Ghosh}, {Giard}, {Gjerl{\o}w}, {Gonz{\'a}lez-Nuevo},
  {G{\'o}rski}, {Gregorio}, {Gruppuso}, {Gudmundsson}, {Hansen}, {Harrison},
  {Hern{\'a}ndez-Monteagudo}, {Herranz}, {Hildebrandt}, {Hobson}, {Hornstrup},
  {Hurier}, {Jaffe}, {Jaffe}, {Jones}, {Juvela}, {Keih{\"a}nen}, {Keskitalo},
  {Kisner}, {Knoche}, {Kunz}, {Kurki-Suonio}, {Lamarre}, {Lasenby}, {Lattanzi},
  {Lawrence}, {Leahy}, {Leonardi}, {Levrier}, {Liguori}, {Lilje},
  {Linden-V{\o}rnle}, {L{\'o}pez-Caniego}, {Lubin}, {Mac{\'{\i}}as-P{\'e}rez},
  {Maggio}, {Maino}, {Mandolesi}, {Mangilli}, {Maris}, {Martin},
  {Mart{\'{\i}}nez-Gonz{\'a}lez}, {Masi}, {Matarrese}, {Melchiorri},
  {Mennella}, {Migliaccio}, {Miville-Desch{\^e}nes}, {Moneti}, {Montier},
  {Morgante}, {Munshi}, {Murphy}, {Naselsky}, {Nati}, {Natoli},
  {N{\o}rgaard-Nielsen}, {Oppermann}, {Orlando}, {Pagano}, {Pajot}, {Paladini},
  {Paoletti}, {Pasian}, {Perotto}, {Pettorino}, {Piacentini}, {Piat},
  {Pierpaoli}, {Plaszczynski}, {Pointecouteau}, {Polenta}, {Ponthieu}, {Pratt},
  {Prunet}, {Puget}, {Rachen}, {Reinecke}, {Remazeilles}, {Renault}, {Renzi},
  {Ristorcelli}, {Rocha}, {Rossetti}, {Roudier}, {Rubi{\~n}o-Mart{\'{\i}}n},
  {Rusholme}, {Sandri}, {Santos}, {Savelainen}, {Scott}, {Spencer},
  {Stolyarov}, {Stompor}, {Strong}, {Sudiwala}, {Sunyaev}, {Suur-Uski},
  {Sygnet}, {Tauber}, {Terenzi}, {Toffolatti}, {Tomasi}, {Tristram}, {Tucci},
  {Valenziano}, {Valiviita}, {Van Tent}, {Vielva}, {Villa}, {Wade}, {Wandelt},
  {Wehus}, {Yvon}, {Zacchei}, \& {Zonca}}]{PIPXLII:2016}
{Planck Collaboration}, {Adam}, R., {Ade}, P.~A.~R., {et~al.}
  2016{\natexlab{b}}, \aap, 596, A103

\bibitem[{{Planck Collaboration} {et~al.}(2016{\natexlab{c}}){Planck
  Collaboration}, {Aghanim}, {Alves}, {Arzoumanian}, {Aumont}, {Baccigalupi},
  {Ballardini}, {Banday}, {Barreiro}, {Bartolo}, {Basak}, {Benabed}, {Bernard},
  {Bersanelli}, {Bielewicz}, {Bonavera}, {Bond}, {Borrill}, {Bouchet},
  {Boulanger}, {Bracco}, {Bucher}, {Burigana}, {Calabrese}, {Cardoso},
  {Chiang}, {Colombo}, {Combet}, {Comis}, {Couchot}, {Coulais}, {Crill},
  {Curto}, {Cuttaia}, {Davis}, {de Bernardis}, {de Rosa}, {de Zotti},
  {Delabrouille}, {Delouis}, {Di Valentino}, {Dickinson}, {Diego}, {Dor{\'e}},
  {Douspis}, {Ducout}, {Dupac}, {Dusini}, {Efstathiou}, {Elsner}, {En{\ss}lin},
  {Eriksen}, {Falgarone}, {Fantaye}, {Ferri{\`e}re}, {Finelli}, {Frailis},
  {Fraisse}, {Franceschi}, {Frolov}, {Galeotta}, {Galli}, {Ganga},
  {G{\'e}nova-Santos}, {Gerbino}, {Ghosh}, {Gonz{\'a}lez-Nuevo}, {G{\'o}rski},
  {Gratton}, {Gregorio}, {Gruppuso}, {Gudmundsson}, {Guillet}, {Hansen},
  {Helou}, {Henrot-Versill{\'e}}, {Herranz}, {Hivon}, {Huang}, {Jaffe},
  {Jaffe}, {Jones}, {Keih{\"a}nen}, {Keskitalo}, {Kisner}, {Krachmalnicoff},
  {Kunz}, {Kurki-Suonio}, {Lagache}, {L{\"a}hteenm{\"a}ki}, {Lamarre},
  {Langer}, {Lasenby}, {Lattanzi}, {Le Jeune}, {Levrier}, {Liguori}, {Lilje},
  {L{\'o}pez-Caniego}, {Lubin}, {Mac{\'{\i}}as-P{\'e}rez}, {Maggio}, {Maino},
  {Mandolesi}, {Mangilli}, {Maris}, {Martin}, {Mart{\'{\i}}nez-Gonz{\'a}lez},
  {Matarrese}, {Mauri}, {McEwen}, {Melchiorri}, {Mennella}, {Migliaccio},
  {Miville-Desch{\^e}nes}, {Molinari}, {Moneti}, {Montier}, {Morgante}, {Moss},
  {Naselsky}, {Natoli}, {Neveu}, {N{\o}rgaard-Nielsen}, {Oppermann},
  {Oxborrow}, {Pagano}, {Paoletti}, {Partridge}, {Perdereau}, {Perotto},
  {Pettorino}, {Piacentini}, {Plaszczynski}, {Polenta}, {Rachen}, {Rebolo},
  {Reinecke}, {Remazeilles}, {Renzi}, {Ristorcelli}, {Rocha}, {Rossetti},
  {Roudier}, {Ruiz-Granados}, {Salvati}, {Sandri}, {Savelainen}, {Scott},
  {Sirignano}, {Soler}, {Suur-Uski}, {Tauber}, {Tavagnacco}, {Tenti},
  {Toffolatti}, {Tomasi}, {Tristram}, {Trombetti}, {Valiviita}, {Vansyngel},
  {Van Tent}, {Vielva}, {Villa}, {Wandelt}, {Wehus}, {Zacchei}, \&
  {Zonca}}]{PIPXLIV:2016}
{Planck Collaboration}, {Aghanim}, N., {Alves}, M.~I.~R., {et~al.}
  2016{\natexlab{c}}, \aap, 596, A105

\bibitem[{{Planck Collaboration ES}(2015)}]{PlanckES:2015}
{Planck Collaboration ES}. 2015,
  \url{http://wiki.cosmos.esa.int/planckpla/index.php/Main_Page} (ESA)

\bibitem[{{Puspitarini} {et~al.}(2014){Puspitarini}, {Lallement}, {Vergely}, \&
  {Snowden}}]{Puspitarini:2014}
{Puspitarini}, L., {Lallement}, R., {Vergely}, J.-L., \& {Snowden}, S.~L. 2014,
  \aap, 566, A13

\bibitem[{{Santos} {et~al.}(2011){Santos}, {Corradi}, \& {Reis}}]{Santos:2011}
{Santos}, F.~P., {Corradi}, W., \& {Reis}, W. 2011, \apj, 728, 104

\bibitem[{{Shelton}(1998)}]{Shelton:1998}
{Shelton}, R.~L. 1998, \apj, 504, 785

\bibitem[{{Sun} \& {Reich}(2010)}]{Sun:2010}
{Sun}, X.-H. \& {Reich}, W. 2010, Research in Astronomy and Astrophysics, 10,
  1287

\bibitem[{{Terral} \& {Ferri{\`e}re}(2017)}]{Terral:2017}
{Terral}, P. \& {Ferri{\`e}re}, K. 2017, \aap, 600, A29

\bibitem[{{Vansyngel} {et~al.}(2017){Vansyngel}, {Boulanger}, {Ghosh},
  {Wandelt}, {Aumont}, {Bracco}, {Levrier}, {Martin}, \&
  {Montier}}]{Vansyngel:2017}
{Vansyngel}, F., {Boulanger}, F., {Ghosh}, T., {et~al.} 2017, \aap, 603, A62

\end{thebibliography}
%-------------------------------------------------------------------
%-------------------------------------------------------------------

\onecolumn{
\begin{appendix}

\section{The magnetic field model}
\label{app}

In this appendix, we describe in more detail some of the steps involved in the derivation of the magnetic field model presented in Sect. \ref{sec1}, in particular, the justification of Eq. (\ref{eq_AppA_A}) and how to go from Eq. (\ref{eq_AppA_A}) to Eq. (\ref{eq_AppA_B_1}). 
Except for a few new physical quantities defined below, all other quantities are as defined in Sect. \ref{sec1}. 

Our magnetic field model relies on the frozen-in approximation, which implies that the magnetic field is coupled to the matter, that is, magnetic field lines follow the matter in its motions. Generally, the frozen-in approximation is given by the induction equation in the limit of zero magnetic diffusion:
\begin{equation}
     \frac{\partial \vec{B}}{\partial t}=\nabla \times (\vec{V} \times \vec{B}),
\label{eq_app_1B}
\end{equation}
where $\vec{B} = \nabla \times \vec{A}$ is the magnetic field, $\vec{A}$ the vector potential, and $\vec{V}$ the velocity field. "De-curling" Eq. (\ref{eq_app_1B}) leads to the evolution equation for $\vec{A}$:
\begin{equation}
     \frac{\partial \vec{A}}{\partial t}=\vec{V} \times (\nabla \times \vec{A}) + \nabla f ,
\label{eq_app_1A}
\end{equation}
with $f$ an arbitrary scalar function. Choosing $f = -\vec{V} \cdot \vec{A}$, we can rewrite Eq. (\ref{eq_app_1A}) in the form 
\begin{equation}
     \frac{d \vec{A}}{d t}= - (\nabla \vec{V}) \cdot \vec{A} .
\label{eq_app_1AA}
\end{equation}
Finally, integrating Eq. (\ref{eq_app_1AA}) in time leads to Eq. (\ref{eq_AppA_A}) (also see \citealt{Parker:1970}).

The present vector potential is obtained by introducing Eq. (\ref{eq_AppA_r0}) (valid for purely radial motions) and Eq. (\ref{eq_AppA_A0}) (valid for a uniform initial magnetic field) into Eq.~(\ref{eq_AppA_A}):
\begin{equation}
\vec{A}({\vec{r}}) = r_{0} \nabla r_{0} \, F(\theta,\phi) + \frac{r_{0}^{2}}{r} \left( T_{\theta}(\theta,\phi) \, \vec{e}_{\theta} + T_{\phi}(\theta,\phi) \, \vec{e}_{\phi} \right),
\label{eq_app_2}
\end{equation}
with 
\begin{align}
\label{eq_app_3}
& F(\theta,\phi)= \sin\theta \left(B_{0y}\cos\theta\cos\phi + B_{0z}\sin\theta\sin\phi\cos\phi + B_{0x}\cos\theta\sin\phi \right) , \nonumber \\
& T_{\theta}(\theta,\phi)=B_{0y}\cos^{2}\theta\cos\phi + B_{0z}\sin\theta\cos\theta\sin\phi\cos\phi - B_{0x}\sin^{2}\theta\sin\phi , \\
& T_{\phi}(\theta,\phi) = -B_{0y}\cos\theta\sin\phi + B_{0z}\sin\theta\cos^{2}\phi \nonumber . 
\end{align}
The final magnetic field then follows from $\vec{B} = \nabla \times \vec{A}$:
\begin{equation}
\vec{B}(\vec{r}) = -r_{0} \nabla r_{0} \times \nabla F + \nabla \times \left[ \frac{r_{0}^{2}}{r} \left( T_{\theta} \, \vec{e}_{\theta} + T_{\phi} \, \vec{e}_{\phi} \right) \right].
\label{eq_app_4}
\end{equation}

Now, assume that all the matter that was initially interior to the current shell has been swept up into an infinitely thin shell presently located at $r = R_{\rm sh}(\theta,\phi)$. Then, for $r \le R_{\rm sh}$, the initial radius can be written as 
\begin{equation}
r_{0}(r,\theta,\phi) = R_{\rm sh}(\theta,\phi) \ h \! \left( r - R_{\rm sh}(\theta,\phi) \right) ,
\label{eq_app_5}
\end{equation}
and its spatial derivatives as
\begin{align}
\label{eq_app_6}
& \frac{\partial r_{0}}{\partial r} = R_{\rm sh}(\theta,\phi) \ \delta \! \left( r - R_{\rm sh}(\theta,\phi) \right), \nonumber \\
& \frac{\partial r_{0}}{\partial \theta} = \left( \frac{r_{0}}{r} - \frac{\partial r_{0}}{\partial r} \right) \frac{\partial R_{\rm sh}}{\partial \theta} , \\
& \frac{\partial r_{0}}{\partial \phi} = \left( \frac{r_{0}}{r} - \frac{\partial r_{0}}{\partial r} \right) \frac{\partial R_{\rm sh}}{\partial \phi} , \nonumber
\end{align}
where $h$ is the step function and $\delta$ the delta function.
Noting that $\partial r_{0}/\partial r \gg r_{0}/r$ in the shell, we introduce Eqs. (\ref{eq_app_3}), (\ref{eq_app_5}), and (\ref{eq_app_6}) into  Eq. (\ref{eq_app_4}) and retain only the leading terms in $\partial r_{0}/\partial r$, whereupon the first term of Eq. (\ref{eq_app_4}) becomes
\begin{equation}
-r_{0}\nabla r_{0} \times \nabla F = \frac{r_{0}}{r} \frac{\partial r_{0}}{\partial r} \left[ \frac{1}{r\sin\theta} \left( \frac{\partial R_{\rm sh}}{\partial \theta}  \frac{\partial F}{\partial \phi} - \frac{\partial R_{\rm sh}}{\partial \phi}  \frac{\partial F}{\partial \theta}\right)  \vec{e}_{r} +  \frac{1}{\sin\theta}\frac{\partial F}{\partial \phi} \, \vec{e}_{\theta} -  \frac{\partial F}{\partial \theta} \, \vec{e}_{\phi} \right],
\label{eq_app_7}
\end{equation} 
the second term becomes
\begin{equation}
\nabla \times \left[ \frac{r_{0}^{2}}{r} \left( T_{\theta} \, \vec{e}_{\theta} + T_{\phi} \, \vec{e}_{\phi} \right) \right]
= 2 \frac{r_{0}}{r} \frac{\partial r_{0}}{\partial r}  \left[ \frac{1}{r\sin\theta} \left(-T_{\phi}\sin\theta \frac{\partial R_{\rm sh}}{\partial \theta} + T_{\theta} \frac{\partial R_{\rm sh}}{\partial \phi} \right) \vec{e}_{r} - T_{\phi} \, \vec{e}_{\theta} + T_{\theta} \, \vec{e}_{\phi} \right],
\label{eq_app_8}
\end{equation}
and their sum gives Eq. (\ref{eq_AppA_B_1}).

\end{appendix}
}
%
%%%%%%%%%%%%%%%%%%%%%%%%%%%%%%%%%%%%%%%%%%%%%%%%%%%%%%%%%%%%
\end{document}